\documentclass[aps,prd,groupedaddress,showpacs,showkeys]{revtex4}
\usepackage{latexsym,epsfig,amsmath,amssymb}
\usepackage{amssymb}
\usepackage{amsmath}
\usepackage{amsfonts}
\usepackage{epsfig}
\usepackage{verbatim}
\newcommand{\eq}{\begin{eqnarray}}
\newcommand{\en}{\end{eqnarray}}

\begin{document}

\title{\Large The effect of multi-channel pion-pion scattering
in decays of the $\Upsilon$-family mesons
}

\author{
Yu.S.~Surovtsev$^{(1)\dag}$, P.~Byd\v{z}ovsk\'y$^{(2)}$, T.~Gutsche$^{(3)}$,
R.~Kami\'nski$^{(4)}$, V.E.~Lyubovitskij$^{(3,5,6)}$, and M.~Nagy$^{(7)}$
\vspace*{1.2\baselineskip}}
\affiliation{
(1) {\it Bogoliubov Laboratory of Theoretical Physics, JINR,141 980 Dubna, Russia
}\\
\vspace*{.1\baselineskip}\\
(2) {\it Nuclear Physics Institute of the AS CR, 25068 \v{R}e\v{z}, Czech Republic}\\
\vspace*{.1\baselineskip}\\
(3) {\it Institut f\"ur Theoretische Physik,  Universit\"at T\"ubingen,
Kepler Center
for Astro and Particle Physics, Auf der Morgenstelle 14,
D--72076 T\"ubingen, Germany}\\
\vspace*{.1\baselineskip}\\
(4) {\it Institute of Nuclear Physics of the PAN, Cracow 31342, Poland}\\
\vspace*{.1\baselineskip}\\
(5) {\it Department of Physics, Tomsk State University, 634050 Tomsk, Russia}\\
\vspace*{.1\baselineskip}\\
(6) {\it
Mathematical Physics Department, Tomsk Polytechnic University, Lenin ave.30,
634050 Tomsk, Russia
}\\
\vspace*{.1\baselineskip}\\
(7) {\it Institute of Physics, SAS, Bratislava 84511, Slovak Republic}
\\}

\date{\today}

\begin{abstract}
The effect of isoscalar S-wave multi-channel pion-pion scattering
($\pi\pi\!\to\!\pi\pi,K\overline{K},\eta\eta$) is considered in the analysis
of data on decays of the $\Upsilon$-meson family --
$\Upsilon(2S)\to\Upsilon(1S)\pi\pi$, $\Upsilon(3S)\to\Upsilon(1S)\pi\pi$
and $\Upsilon(3S)\to\Upsilon(2S)\pi\pi$. The analysis, which aims at
studying the scalar mesons, is performed jointly considering the multi-channel pion-pion scattering, which is described in our model-independent approach
based on analyticity and unitarity and using an uniformizing variable method,
and the charmonium decay processes $J/\psi\to\phi(\pi\pi, K\overline{K})$, $\psi(2S)\to J/\psi(\pi\pi)$.
Results of the analysis confirm all our earlier conclusions on the scalar
mesons.
It is also shown that in the final states of the $\Upsilon$-meson family
decays (except for the $\pi\pi$ scattering) the contribution of the coupled
processes, e.g., $K\overline{K}\to\pi\pi$, is important even if these processes
are energetically forbidden.
This is in accordance with our previous conclusions on the wide resonances:
If a wide resonance cannot decay into a channel which opens above its mass
but the resonance is strongly connected with this channel (e.g. the $f_0(500)$
and the $K\overline{K}$ channel), one should consider this resonance as a
multi-channel state with allowing for the indicated channel taking into account
the Riemann-surface sheets related to the threshold branch-point of this channel
and performing the combined analysis of the considered and coupled channels.

\end{abstract}



\maketitle

\section{Introduction}

In the analysis of data on decays of the $\Upsilon$-meson family --$\Upsilon(2S)\to\Upsilon(1S)\pi\pi$, $\Upsilon(3S)\to\Upsilon(1S)\pi\pi$ and $\Upsilon(3S)\to\Upsilon(2S)\pi\pi$ -- the contribution of multi-channel $\pi\pi$ scattering in the final-state interactions is considered.
The analysis, which aims at studying the scalar mesons, is performed jointly considering the isoscalar S-wave processes $\pi\pi\!\to\!\pi\pi,K\overline{K},\eta\eta$, which are described in our model-independent approach based on analyticity and unitarity and using an uniformization procedure, and the charmonium decay processes
$J/\psi\to\phi(\pi\pi, K\overline{K})$, $\psi(2S)\to J/\psi(\pi\pi)$.
\vspace*{0.5mm}

Importance of studying properties of scalar mesons is related to the obvious fact that a comprehension of these states is necessary in principle for the most profound topics concerning the QCD vacuum, because these sectors affect each other especially strongly due to possible "direct" transitions between them. However the problem of interpretation of the scalar mesons is faraway to be solved completely \cite{PDG-12}.

E.g., applying our model-independent method in the 3-channel analyses of processes $\pi\pi\to\pi\pi,K\overline{K},\eta\eta,\eta\eta^\prime$ \cite{SBKN-PRD10,SBL-prd12} we have obtained parameters of the $f_0(500)$ and $f_0(1500)$ which differ considerably from results of analyses which utilize other methods (mainly those based on dispersion relations and Breit--Wigner approaches).

To make our approach more convincing, to confirm obtained results and to diminish inherent arbitrariness, we have utilized the derived model-independent amplitudes for multi-channel $\pi\pi$ scattering calculating the contribution of final-state interactions in decays $J/\psi\to\phi(\pi\pi, K\overline{K})$, $\psi(2S)\to J/\psi(\pi\pi)$ and $\Upsilon(2S)\to\Upsilon(1S)\pi\pi$ \cite{SBGLKN-npbps13,SBLKN-prd14}.

Here we add to the analysis the data on decays $\Upsilon(3S)\to\Upsilon(1S)\pi\pi$ and $\Upsilon(3S)\to\Upsilon(2S)\pi\pi$ from CLEO(94) Collaboration. A distinction of the $\Upsilon(3S)$ decays from the above ones consists in the fact that in this case a phase space cuts off, as if, possible contributions which might interfere destructively with the $\pi\pi$-scattering contribution giving a characteristic 2-humped shape of the energy dependence of di-pion spectrum in decay $\Upsilon(3S)\to\Upsilon(1S)\pi\pi$.

After establishing the 2-humped shape of di-pion spectrum Lipkin and Tuan \cite{Lipkin-Tuan} have suggested that the decay $\Upsilon(3S)\to\Upsilon(1S)\pi\pi$ proceeds as follows:
$~~~\Upsilon(3S)\to B^*\overline{B}^*\to B^*\overline{B}\pi\to B\overline{B}\pi\pi\to\Upsilon(1S)\pi\pi$.\\ Then in the heavy-quarkonium limit,
when neglecting recoil of the final-quarkonium state, they obtained that the amplitude contains a term proportional to ${{\bf p}_1\!*{\bf p}_2}\propto\cos\theta_{12}$ ($\theta_{12}$ is the angle between the pion three-momenta ${\bf p}_1$  and ${\bf p}_2$) multiplied by some function of the kinematic invariants. If the latter were a constant, then the distribution
$d\Gamma/d\cos\theta_{12}\propto\cos\theta_{12}^2$ (and $d\Gamma/d M_{\pi\pi}$) would have the 2-humped shape. However, this scenario was not tested numerically by fitting to data. It is possible that this
effect is negligible due to the small coupling of the $\Upsilon$ to the b-flavored sector.

In his work \cite{Moxhay}, Moxhay has suggested that the 2-humped shape is a result of interference
between two parts of the decay amplitude.
One part, in which the $\pi\pi$ final state interaction is allowed for, is related to a mechanism
which acts well in the decays $\psi(2S)\to J/\psi(\pi\pi)$ and $\Upsilon(2S)\to\Upsilon(1S)\pi\pi$
and which, obviously, should operate also in the process $\Upsilon(3S)\to\Upsilon(1S)\pi\pi$.
The other part is responsible for the Lipkin -- Tuan mechanism. Though there remains nothing from
the latter because the author says that the term containing ${\bf p}_1\!*{\bf p}_2$ does not dominate
this part of amplitude and ``the other tensor structures conspire to give a distribution in
$M_{\pi\pi}$ that is more or less flat'' -- indeed, constant.

It seems, the approach of work \cite{Komada-Ishida-Ishida} resembles the above one. The authors
have supposed simply that a pion pair is formed in the $\Upsilon(3S)$ decay both as a result of
re-scattering and ``directly". One can, however, believe that the latter is not reasonable because
the pions interact strongly, inevitably.

We show that the indicated effect of destructive interference can be achieved taking into account our previous conclusions on the wide resonances \cite{SBLKN-prd14,SBLKN-jpgnpp14}, without the doubtful assumptions.

\section{The model-independent amplitudes for multi-channel $\pi\pi$ scattering}

Considering the multi-channel $\pi\pi$ scattering, we shall deal with the 3-channel case (namely with $\pi\pi\!\to\!\pi\pi,K\overline{K},\eta\eta$) because it was shown \cite{SBLKN-jpgnpp14,SBKLN-PRD12} that this is a minimal number of coupled channels needed for obtaining correct values of scalar-isoscalar resonance parameters.

\vspace*{0.5mm}
\begin{itemize}
\item{\underline{Resonance representations on the 8-sheeted Riemann surface}}
\end{itemize}
\vspace*{0.6mm}
The 3-channel $S$-matrix is determined on the 8-sheeted Riemann surface. The matrix elements $S_{ij}$, where $i,j=1,2,3$ denote channels, have the right-hand cuts along the real axis of the $s$ complex plane ($s$ is the invariant total energy squared), starting with the channel thresholds $s_i$ ($i=1,2,3$), and the left-hand cuts related to the crossed channels.
The Riemann-surface sheets are numbered according to the signs of analytic continuations of the square roots $\sqrt{s-s_i}~~(i=1,2,3)$ as follows:\\
\begin{center}
\begin{tabular}{|c|c|c|c|c|c|c|c|c|} \hline
{} & ~I~ & ~II~ & ~III~ & ~IV~ & ~V~ & ~VI~ & ~VII~ & ~VIII~ \\ \hline
{~$\mbox{Im}\sqrt{s-s_1}$~} & $+$ & $-$ & $-$ & $+$ & $+$ & $-$ & $-$ & $+$ \\
{~$\mbox{Im}\sqrt{s-s_2}$~} & $+$ & $+$ & $-$ & $-$ & $-$ & $-$ & $+$ & $+$\\
{~$\mbox{Im}\sqrt{s-s_3}$~} & $+$ & $+$ & $+$ & $+$ & $-$ & $-$ & $-$ & $-$\\
\hline
\end{tabular}
\end{center}
\vspace*{0.1mm}
An adequate allowance for the Riemann surface structure is performed taking the following uniformizing variable
\cite{SBL-prd12} where we have neglected the $\pi\pi$-threshold branch-point and taken into account the $K\overline{K}$- and $\eta\eta$-threshold branch-points and the left-hand branch-point at $s=0$ related to the crossed channels:
\begin{equation}\label{w}
w=\frac{\sqrt{(s-s_2)s_3} + \sqrt{(s-s_3)s_2}}{\sqrt{s(s_3-s_2)}}~~~~(s_2=4m_K^2 ~ {\rm and}~ s_3=4m_\eta^2)
\end{equation}
(reasons and substantiation for neglecting the $\pi\pi$-threshold branch-point can be found in works \cite{SBL-prd12,KMS-96}).

Resonance representations on the Riemann surface are obtained using formulas from \cite{SBL-prd12,KMS-96}, expressing analytic continuations of the $S$-matrix elements to all sheets in terms of those on the physical (I) sheet that have only the resonances zeros (beyond the real axis), at least, around the physical region.
In the 3-channel case, there are {\it 7 types} of resonances corresponding to 7 possible situations when there are resonance zeros on sheet I only in $S_{11}$ -- ({\bf a}); ~~$S_{22}$ -- ({\bf b}); ~~$S_{33}$ -- ({\bf c}); ~~$S_{11}$ and $S_{22}$ -- ({\bf d}); ~~$S_{22}$ and $S_{33}$ -- ({\bf e}); ~~$S_{11}$ and $S_{33}$ -- ({\bf f}); ~~$S_{11}$, $S_{22}$ and $S_{33}$ -- ({\bf g}). The resonance of every type is represented by the pair of complex-conjugate \underline{clusters} (of poles and zeros on the Riemann surface).

The variable $w$ eq.(\ref{w}) maps a model of the 8-sheeted Riemann surface, allowing for the neglect
of the $\pi\pi$-threshold branch-point, onto the $w$-plane divided into two parts by
a unit circle centered at the origin (Fig. 1).
\begin{figure}[!htb]
\begin{center}
\includegraphics[width=0.36\textwidth,angle=-90]{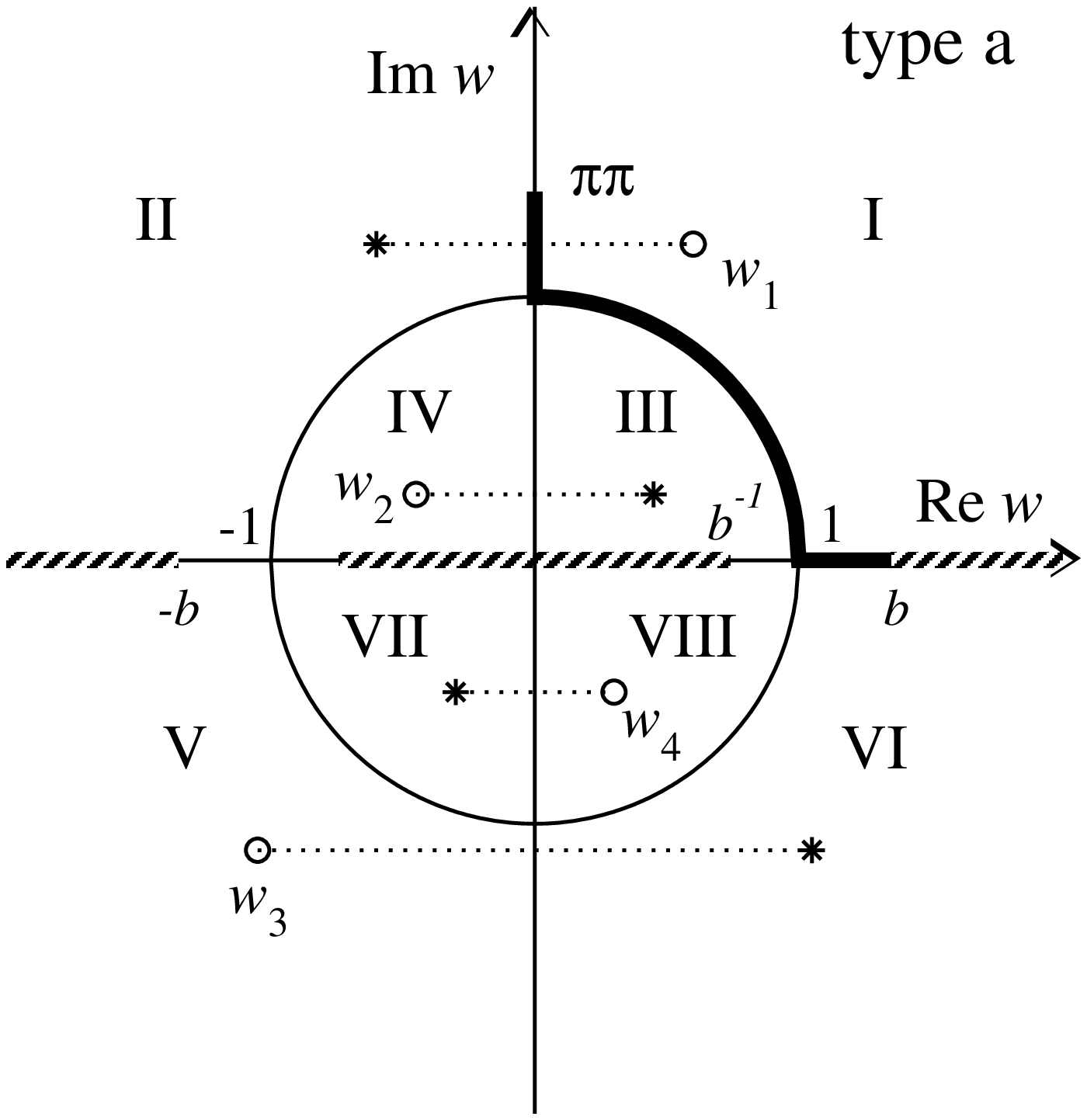}
\hspace*{-0.1cm}
\includegraphics[width=0.36\textwidth,angle=-90]{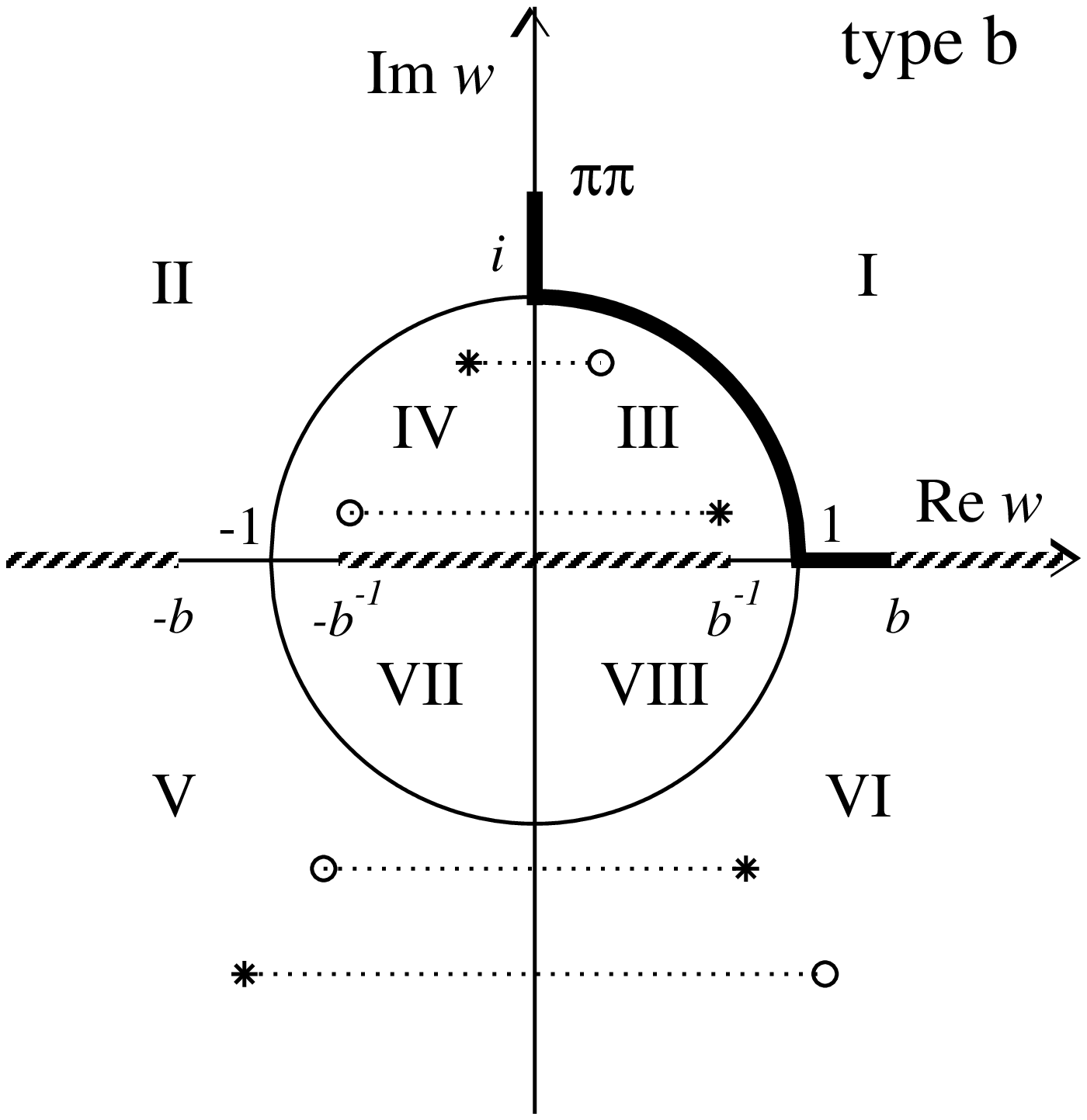}\\
\vspace*{0.5cm}
\includegraphics[width=0.36\textwidth,angle=-90]{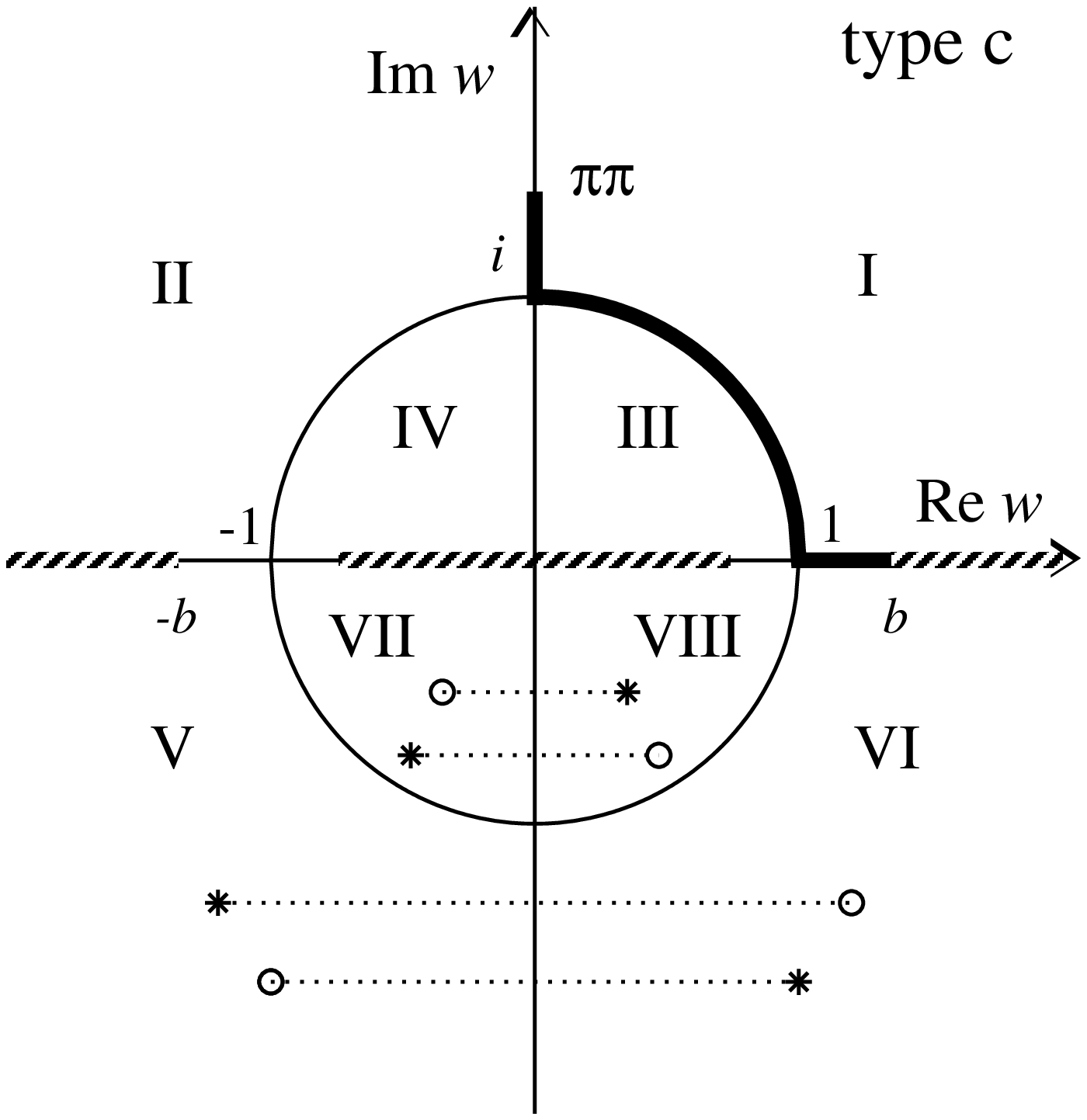}
\hspace*{-0.1cm}
\includegraphics[width=0.36\textwidth,angle=-90]{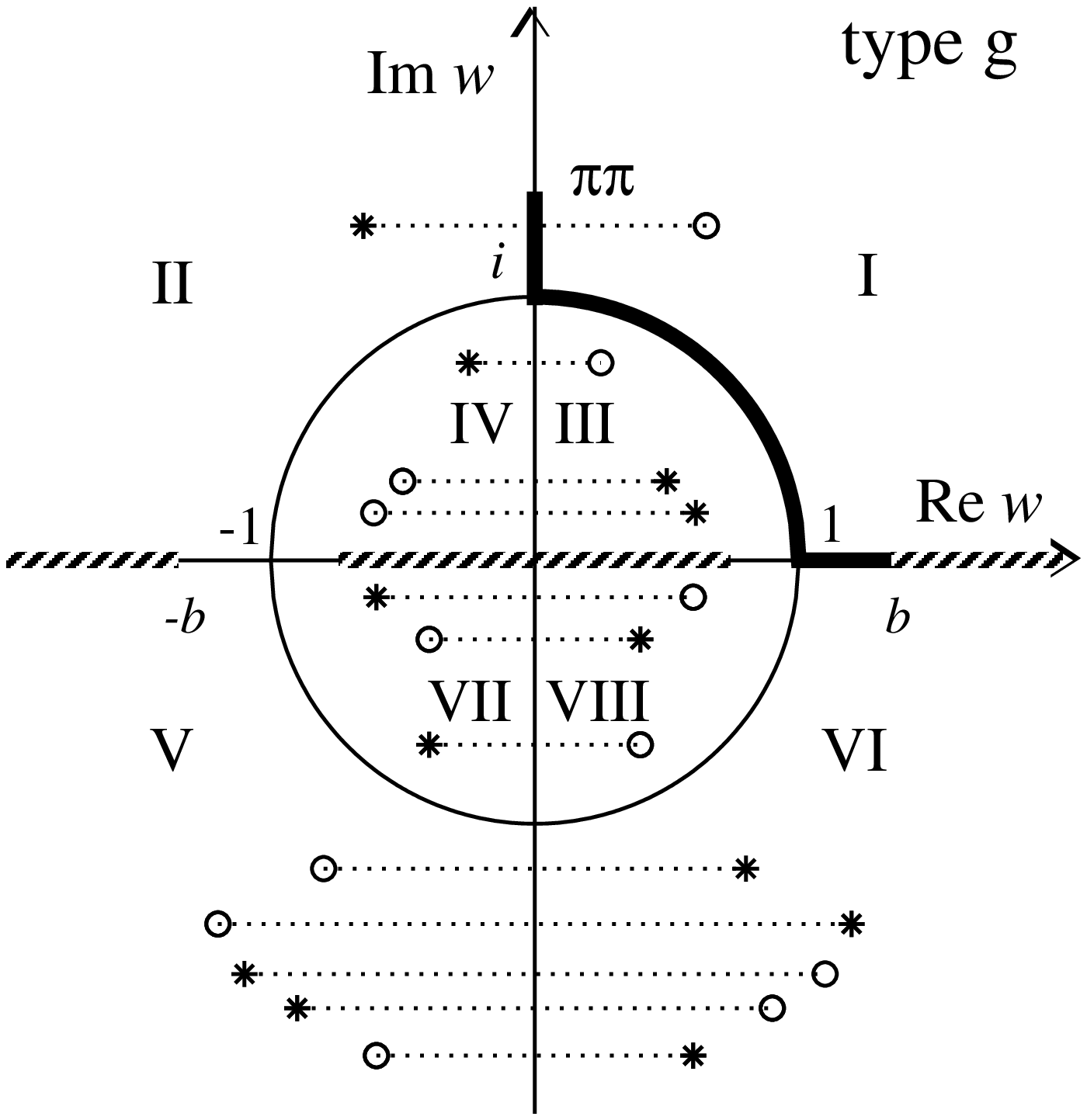}
\vspace*{0.3cm}
\caption{Uniformization $w$-plane: Representation of resonances of
types ({\bf a}), ({\bf b}), ({\bf c}) and ({\bf g}) in the
3-channel $\pi\pi$-scattering $S$-matrix element.}
\end{center}\label{fig:lw_plane}
\end{figure}
The semi-sheets I (III), II (IV), V (VII) and VI (VIII) are mapped onto the exterior (interior) of the unit disk in the 1st, 2nd, 3rd and 4th quadrants, respectively. The physical region, shown by the thick line, extends from the point $\pi\pi$ on the imaginary axis (the first $\pi\pi$ threshold corresponding to $s_1$) along this axis down to the point {\it i} on the unit circle (the second threshold corresponding to $s_2$). Then it extends further along the unit circle
clockwise in the 1st quadrant to point 1 on the real axis (the third threshold
corresponding to $s_3$) and then along the real axis to the point
$b=(\sqrt{s_2}+\sqrt{s_3})/\sqrt{s_3-s_2}$ into which $s=\infty$ is mapped on the $w$-plane. The intervals $(-\infty,-b]$, $[-b^{-1},b^{-1}]$, $[b,\infty)$ on the real axis are the images of the corresponding edges of the left-hand cut of the $\pi\pi$-scattering amplitude. In Fig. 1, the 3-channel resonances of the types ({\bf a}), ({\bf b}), ({\bf c}) and ({\bf g}), met in the analysis, in $S_{11}(w)$ are represented  by the poles ($*$) and zeroes ($\circ$) symmetric to these poles with respect to the imaginary axis giving corresponding pole clusters. The ``pole--zero'' symmetry guarantees the elastic unitarity of $\pi\pi$ scattering in the ($\pi\pi$, $i$) interval.

\newpage
\begin{itemize}
\item{\underline{The $S$-matrix parametrization}}
\end{itemize}
\vspace*{0.1cm}
The $S$-matrix elements $S_{ij}$ are parameterized using the Le Couteur-Newton relations \cite{LeCou}. On the $w$-plane, we have derived for them:
\begin{equation}\label{w:LeCouteur-Newton}
S_{11}=\frac{d^* (-w^*)}{d(w)},~~~~~~~~
S_{22}=\frac{d(-w^{-1})}{d(w)},~~~~~~~~
S_{33}=\frac{d(w^{-1})}{d(w)},
\end{equation}
\begin{equation}
S_{11}S_{22}-S_{12}^2=\frac{d^*({w^*}^{-1})}{d(w)},~~~~~~~~
S_{11}S_{33}-S_{13}^2=\frac{d^*(-{w^*}^{-1})}{d(w)}.~~
\end{equation}
The $d(w)$ is the Jost matrix determinant.
The 3-channel unitarity requires the following relations to hold for physical $w$-values:
\begin{equation}
|d(-w^*)|\leq |d(w)|,\quad |d(-w^{-1})|\leq |d(w)|,\quad |d(w^{-1})|\leq
|d(w)|,
\end{equation}
\begin{equation}
|d({w^*}^{-1})|=|d(-{w^*}^{-1})|=|d(-w)|=|d(w)|.
\end{equation}

The $S$-matrix elements in Le Couteur--Newton relations (\ref{w:LeCouteur-Newton}) are taken as the products~~$S=S_B S_{res}$; the main (\underline{model-independent}) contribution of resonances, given by the pole clusters, is included in the resonance part $S_{res}$; possible remaining small (\underline{model-dependent}) contributions of resonances and influence of channels which are not taken explicitly into account in the uniformizing variable are included in the background part $S_B$. The d-function is:\\ for the resonance part
\begin{equation}
d_{res}(w)=w^{-\frac{M}{2}}\prod_{r=1}^{M}(w+w_{r}^*)~~~(M~ \mbox{is the number of resonance zeros}),
\end{equation}
for the background part
\begin{equation}
d_B=\mbox{exp}[-i\sum_{n=1}^{3}\frac{\sqrt{s-s_n}}{2m_n}(\alpha_n+i\beta_n)]
\end{equation}
where
$$\alpha_n=a_{n1}+a_{n\sigma}\frac{s-s_\sigma}{s_\sigma}\theta(s-s_\sigma)+
a_{nv}\frac{s-s_v}{s_v}\theta(s-s_v)
$$
$$\beta_n=b_{n1}+b_{n\sigma}\frac{s-s_\sigma}{s_\sigma}\theta(s-s_\sigma)+
b_{nv}\frac{s-s_v}{s_v}\theta(s-s_v)$$
with $s_\sigma$ the $\sigma\sigma$ threshold, $s_v$ the combined threshold of the
$\eta\eta^{\prime},~\rho\rho,~\omega\omega$ channels.
The resonance zeros $w_{r}$ and the background parameters were fixed by fitting to data on processes ~$\pi\pi\to\pi\pi,K\overline{K},\eta\eta$ with adding the data on decays $J/\psi\to\phi(\pi\pi, K\overline{K})$, $\psi(2S)\to J/\psi(\pi\pi)$ from the Crystal Ball, DM2, Mark~II, Mark~III, and BES~II Collaborations.

For the data on multi-channel $\pi\pi$ scattering we used the results of phase
analyses which are given for phase shifts of the amplitudes $\delta_{\alpha\beta}$
and for the modules of the $S$-matrix elements
$\eta_{\alpha\beta}=|S_{\alpha\beta}|$ ($\alpha,\beta=1,~2,~3$):
\begin{equation}
S_{\alpha\alpha}=\eta_{\alpha\alpha}e^{2i\delta_{\alpha\alpha}},~~~~~
S_{\alpha\beta}=i\eta_{\alpha\beta}e^{i\phi_{\alpha\beta}}.
\end{equation}
If below the third threshold there is the 2-channel unitarity then
the relations
\begin{equation}
\eta_{11}=\eta_{22}, ~~ \eta_{12}=(1-{\eta_{11}}^2)^{1/2},~~
\phi_{12}=\delta_{11}+\delta_{22}
\end{equation}
are fulfilled in this energy region.

For the $\pi\pi$ scattering, the data from the threshold to 1.89~GeV are taken from many works
\cite{pipi-data}. For $\pi\pi\to K\overline{K}$, practically all the accessible data are used
\cite{pipiKK-data}. For $\pi\pi\to\eta\eta$, we have taken the data for $|S_{13}|^2$ from the
threshold to 1.72~GeV \cite{pipi_eta_eta-data}.

We have found a following more preferable scenarios: the $f_0(500)$ is described by the cluster of type ({\bf a}); the $f_0(1370)$ and $f_0(1500)$, type ({\bf c}) and $f_0^\prime(1500)$, type ({\bf g}); the $f_0(980)$ is represented only by the pole on sheet~II and shifted pole on sheet~III. However, the $f_0(1710)$ can be described by clusters either of type ({\bf b}) or ({\bf c}). For definiteness, we have taken type~({\bf c}).

Analyzing these data, we have obtained two solutions which are distinguished mainly in the width of $f_0(500)$. Further we show the solution which has survived after adding to the analysis the data on decays $J/\psi\to\phi(\pi\pi, K\overline{K})$ from the Mark~III, DM2 and BES~II Collaborations.

A comparison of the description with the experimental data on multi-channel $\pi\pi$ scattering  is shown in Fig.~2.
%\vspace{4mm}

\begin{figure}[!thb]
\begin{center}
\includegraphics[width=0.46\textwidth,angle=0]{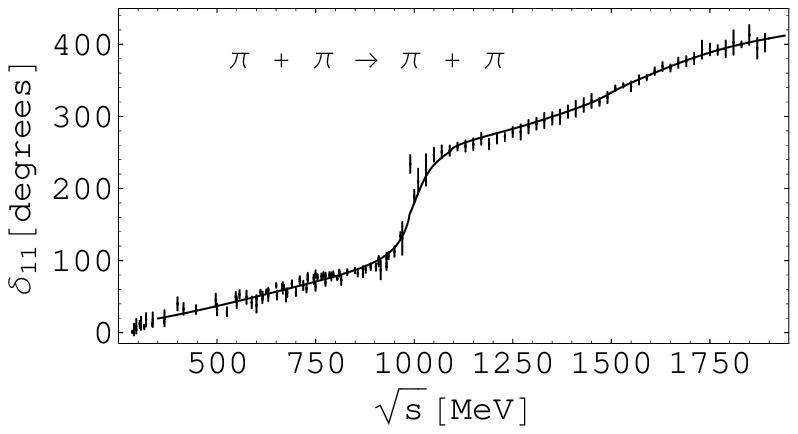}
\includegraphics[width=0.46\textwidth,angle=0]{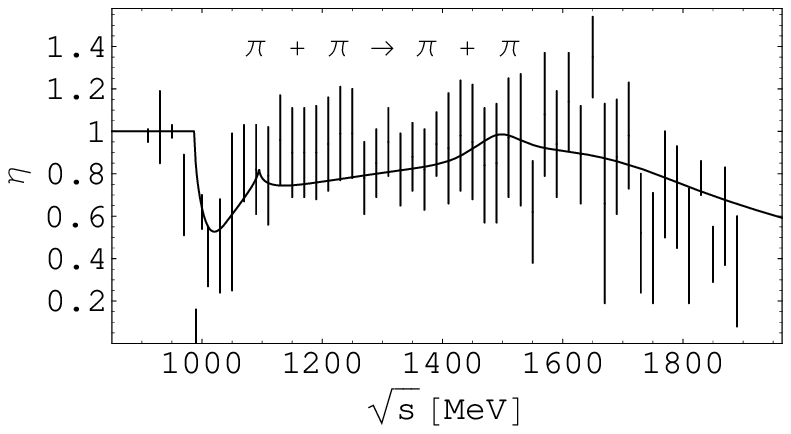}\\
\includegraphics[width=0.46\textwidth,angle=0]{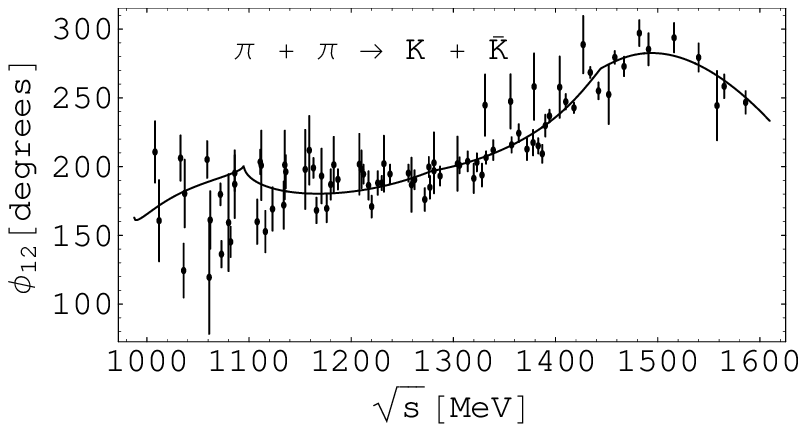}
\includegraphics[width=0.46\textwidth,angle=0]{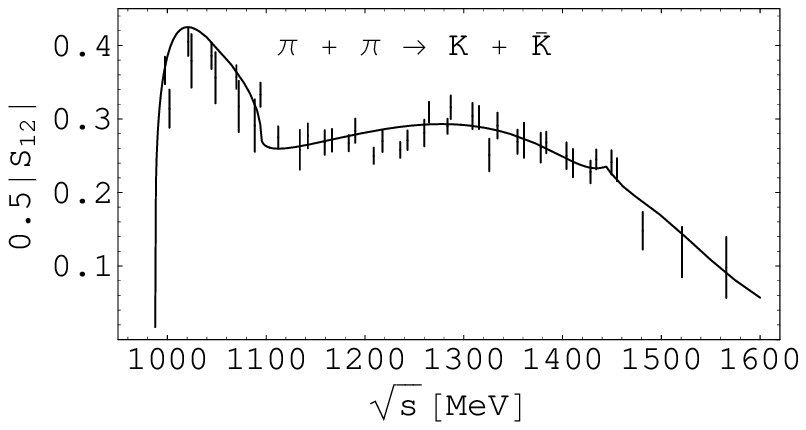}\\
\vspace*{-0.0cm}                                                                      \includegraphics[width=0.48\textwidth,angle=0]{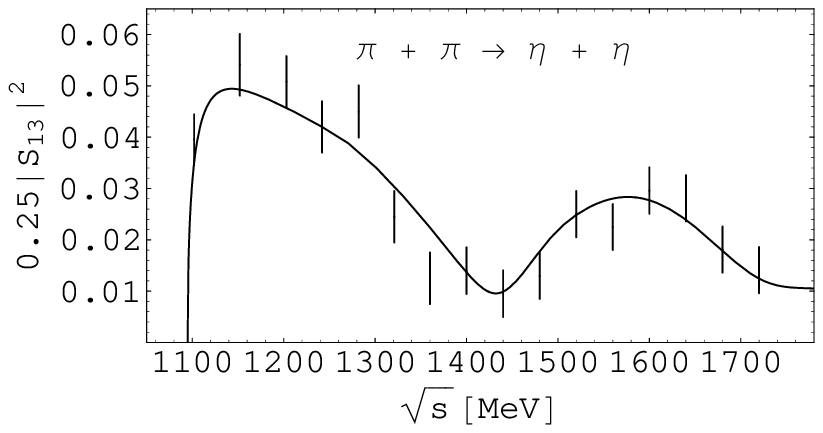}
\vskip -.3cm
\caption{The phase shifts and modules of the $S$-matrix element in the S-wave $\pi\pi$-scattering (upper panel), in $\pi\pi\to K\overline{K}$ (middle panel), and the squared module of the $\pi\pi\to\eta\eta$ $S$-matrix element (lower figure).}
\end{center}\label{fig:fitting}
\end{figure}

In Table~\ref{tab:clusters} we show the obtained pole clusters for the resonances on the complex energy plane $\sqrt{s}$. The poles on sheets III, V, and VII and VI, corresponding to the $f_0^\prime(1500)$, are of the second and third order, respectively (this is an approximation).
\begin{table}[!htb]
\caption{The pole clusters for $f_0$ resonances on the $\sqrt{s}$-plane.
~$\sqrt{s_r}\!=\!{\rm E}_r\!-\!i\Gamma_r/2$ in MeV.}
\label{tab:clusters}
\vspace*{0.05cm}
\def\arraystretch{1.5}
\begin{tabular}{|c|c|c|c|c|c|c|c|}
\hline ${\rm Sheet}$ & {} & $f_0(500)$ & $f_0(980)$ & $f_0(1370)$ & $f_0(1500)$ & $f_0^\prime(1500)$ & $f_0(1710)$ \\ \hline
II & {${\rm E}_r$} & $514.5\pm12.4$ & $1008.1\pm3.1$\! & {} & {} & $1512.7\pm4.9$ & {} \\
{} & {$\Gamma_r/2$} & $465.6\pm5.9$ & $32.0\pm1.5$ & {} & {} & $285.8\pm12.9$ & {} \\
\hline III & {${\rm E}_r$} & $544.8\pm17.7$ & $976.2\pm5.8$ & $1387.6\pm24.4$ & {} & $1506.2\pm9.0$ & {} \\{} & {$\Gamma_r/2$} & $465.6\pm5.9$ & $53.0\pm2.6$ & $166.9\pm41.8$ & {} & \!\!\!$127.9\pm10.6$ & {} \\
\hline IV & {${\rm E}_r$} & {} & {} & 1387.6$\pm$24.4 & {} & 1512.7$\pm$4.9 & {} \\
{} & {$\Gamma_r/2$} & {} & {} & $178.5\pm37.2$ & {} & $216.0\pm17.6$ & {} \\
\hline V & {${\rm E}_r$} & {} & {} & 1387.6$\pm$24.4 & $1493.9\pm3.1$ & $1498.9\pm7.2$ & $1732.8\pm43.2$ \\
{} & {$\Gamma_r/2$} & {} & {} & $260.9\pm73.7$ & $72.8\pm3.9$ & $142.2\pm6.0$ & $114.8\pm61.5$ \\
\hline VI & {${\rm E}_r$} & $566.5\pm29.1$ & {} & 1387.6$\pm$24.4 & $1493.9\pm5.6$
& $1511.4\pm4.3$ & 1732.8$\pm$43.2 \\
{} & {$\Gamma_r/2$} & $465.6\pm5.9$ & {} & $249.3\pm83.1$ & $58.4\pm2.8$ & $179.1\pm4.0$ & $111.2\pm8.8$ \\
\hline VII & {${\rm E}_r$} & $536.2\pm25.5$ & {} & {} & $1493.9\pm5.0$ & $1500.5\pm9.3$ & 1732.8$\pm$43.2 \\
{} & {$\Gamma_r/2$} & $465.6\pm5.9$ & {} & {} & $47.8\pm9.3$ & $99.7\pm18.0$ & $55.2\pm38.0$ \\
\hline VIII & {${\rm E}_r$} & {} & {} & {} & $1493.9\pm3.2$ & 1512.7$\pm$4.9 & 1732.8$\pm$43.2 \\
{} & {$\Gamma_r/2$} & {} & {} & {} & $62.2\pm9.2$ & $299.6\pm14.5$ & $58.8\pm16.4$ \\
\hline
\end{tabular}
\end{table}

The obtained background parameters are:
\underline{$a_{11}=0.0$, $a_{1\sigma}=0.0199$, $a_{1v}=0.0$,} \underline{$b_{11}=b_{1\sigma}=0.0$, $b_{1v}=0.0338$,} $a_{21}=-2.4649$, $a_{2\sigma}=-2.3222$, $a_{2v}=-6.611$, $b_{21}=b_{2\sigma}=0.0$, $b_{2v}=7.073$, $b_{31}=0.6421$, $b_{3\sigma}=0.4851$, $b_{3v}=0$; $s_\sigma=1.6338~{\rm GeV}^2$, $s_v=2.0857~{\rm GeV}^2$.

The very simple description of the $\pi\pi$-scattering background (underlined numbers) confirms well our assumption \underline{$S=S_B S_{res}$} and also that representation of multi-channel resonances by the pole clusters on the uniformization plane is good and quite sufficient.

\underline{It is important that we have obtained practically zero background of the}
\underline{$\pi\pi$ scattering in the scalar-isoscalar channel} because a reasonable and simple description of the background should be a criterion for the correctness of the approach. Furthermore, this shows that the consideration of the left-hand branch-point at $s=0$ in the uniformizing variable solves partly a problem of some approaches (see, e.g., \cite{Achasov94}) that the wide-resonance parameters are strongly controlled by the non-resonant background.

Generally, {\it wide multi-channel states are most adequately represented by pole clusters}, because the pole clusters give the main model-independent effect of resonances. The pole positions are rather stable characteristics for various models, whereas masses and widths are very model-dependent for wide resonances.

However, mass values are needed in some cases, e.g., in mass relations for multiplets. Therefore, we stress that such parameters of the wide multi-channel states, as {\it masses, total widths and coupling constants with channels, should be calculated using the poles on sheets II, IV and VIII}, because only on these sheets the analytic continuations have the forms: $$\propto 1/S_{11}^{\rm I},~~
\propto 1/S_{22}^{\rm I}~~{\rm and}~~\propto 1/S_{33}^{\rm I},$$
respectively, i.e., the pole positions of resonances are at the same points of the complex-energy plane, as the resonance zeros on the physical sheet, and are not shifted due to the coupling of channels. E.g., if the resonance part of amplitude is taken as
\begin{equation}
T^{res}=\sqrt{s}~\Gamma_{el}/(m_{res}^2-s-i\sqrt{s}~\Gamma_{tot}),
\end{equation}
for the mass and total width, one obtains
\begin{equation}
m_{res}=\sqrt{{\rm E}_r^2+\left(\Gamma_r/2\right)^2}~~~
{\rm and}~~~\Gamma_{tot}=\Gamma_r,
\end{equation}
where the pole position $\sqrt{s_r}\!=\!{\rm E}_r\!-\!i\Gamma_r/2$ must be taken
on sheets II, IV, VIII, depending on the resonance classification.

In Table~\ref{tab:mass-width} we show the obtained values of masses and total widths of the $f_0$ resonances.
\begin{table}[htb!]
\caption{\large The masses and total widths of the $f_0$ resonances.}
\vspace*{0.05cm}
\label{tab:mass-width}
{%\small
%\scriptsize
%\normalsize
\def\arraystretch{1.5}
\begin{tabular}{|c|c|c|c|c|c|c|}
\hline {} & $f_0(500)$ & $f_0(980)$ & $f_0(1370)$ & $f_0(1500)$ & $f_0^\prime(1500)$ & $f_0(1710)$\\ \hline
$m_{res}$[MeV] & 693.9$\pm$10.0 & 1008.1$\pm$3.1 & 1399.0$\pm$24.7 & 1495.2$\pm$3.2 & 1539.5$\pm$5.4 & 1733.8$\pm$43.2 \\ \hline
$\Gamma_{tot}$[MeV] & 931.2$\pm$11.8 & 64.0$\pm$3.0
& 357.0$\pm$74.4 & 124.4$\pm$18.4 & 571.6$\pm$25.8 & 117.6$\pm$32.8 \\
\hline
\end{tabular}}
\end{table}

\section{The contribution of multi-channel $\pi\pi$ scattering in the final states of decays of $\psi$- and $\Upsilon$-meson families}

For decays $J/\psi\to\phi\pi\pi,\phi K\overline{K}$ we have taken data from Mark III \cite{MarkIII}, DM2 \cite{DM2} and BES II \cite{BES} Collaborations;
for $\psi(2S)\to J/\psi(\pi^+\pi^-)$ from Mark~II \cite{Mark_II}; for $\psi(2S)\to J/\psi(\pi^0\pi^0)$ from Crystal Ball(80) \cite{Crystal_Ball(80)}; for $\Upsilon(2S)\to\Upsilon(1S)(\pi^+\pi^-,\pi^0\pi^0)$ from Argus \cite{Argus}, CLEO \cite{CLEO}, CUSB \cite{CUSB}, and Crystal Ball(85) \cite{Crystal_Ball(85)} Collaborations; finally for $\Upsilon(3S)\to\Upsilon(1S)(\pi^+\pi^-,\pi^0\pi^0)$ and $\Upsilon(3S)\to\Upsilon(2S)(\pi^+\pi^-,\pi^0\pi^0)$ from CLEO(94) Collaboration \cite{CLEO(94)}.

Formalism for calculating di-meson mass distributions of decays $J/\psi\to\phi(\pi\pi, K\overline{K})$ and $V^{\prime}\to V\pi\pi$ ($V=\psi,\Upsilon$) can be found in Ref. \cite{MP-prd93}. There is assumed that pairs of pseudo-scalar mesons of final states have $I=J=0$ and only they undergo strong interactions, whereas a final vector meson ($\phi$, $V$) acts as a spectator. The amplitudes for decays are related with the scattering amplitudes $T_{ij}$ $(i,j=1-\pi\pi,2-K\overline{K})$ as follows
\begin{equation}F(J/\psi\to\phi\pi\pi)=\sqrt{2/3}~[c_1(s)T_{11}+c_2(s)T_{21}],\end{equation}
\begin{equation}F(J/\psi\to\phi K\overline{K})=\sqrt{1/2}~[c_1(s)T_{12}+c_2(s)T_{22}],\end{equation}
\begin{equation}
F(V(2S)\to V(1S)\pi\pi~(V=\psi,\Upsilon))=[(d_1,e_1)T_{11}+(d_2,e_2)T_{21}],\end{equation}
\begin{equation}
F(\Upsilon(3S)\to \Upsilon(1S,2S)\pi\pi)=[(f_1,g_1)T_{11}+(f_2,g_2)T_{21}]\end{equation}
where  ~~$c_1=\gamma_{10}+\gamma_{11}s$,  ~~$c_2=\alpha_2/(s-\beta_2)+\gamma_{20}+\gamma_{21}s$, ~~$(d_i,e_i)=(\delta_{i0},\rho_{i0})+(\delta_{i1},\rho_{i1})s$ and ~~$(f_i,g_i)=(\omega_{i0},\tau_{i0})+(\omega_{i1},\tau_{i1})s$~~ are functions of couplings of the $J/\psi$, $\psi(2S)$, $\Upsilon(2S)$ and $\Upsilon(3S)$ to channel~$i$; ~$\alpha_2$, $\beta_2$, $\gamma_{i0}$, $\gamma_{i1}$, $\delta_{i0}$, $\rho_{i0}$, $\delta_{i1}$, $\rho_{i1}$, $\omega_{i0}$, $\omega_{i1}$, $\tau_{i0}$ and $\tau_{i1}$ are free parameters. The pole term in $c_2$ is an approximation of possible $\phi K$ states, not forbidden by OZI rules when considering quark diagrams of these processes. Obviously this pole should be situated on the real $s$-axis below the $\pi\pi$ threshold.

The expressions for decays $J/\psi\to\phi(\pi\pi, K\overline{K})$
\begin{equation}
N|F|^{2}\sqrt{(s-s_i)[m_\psi^{2}-(\sqrt{s}-m_\phi)^{2}][m_\psi^2-(\sqrt{s}+m_\phi)^2]}
\end{equation}
and the analogues relations for $V(2S)\to V(1S)\pi\pi~(V=\psi,\Upsilon)$ and $\Upsilon(3S)\to \Upsilon(1S,2S)\pi\pi$ give the di-meson mass distributions. N (normalization to experiment) is 0.7512 for Mark~III, 0.3705 for DM2, 5.699 for BES~II, 1.015 for Mark~II, 0.98 for Crystal Ball(80), 4.3439 for Argus, 2.1776 for CLEO, 1.2011 for CUSB, 0.0788 for Crystal Ball(85), and, finally, for CLEO(94): 0.5096 and 0.2235 for $\Upsilon(3S)\to\Upsilon(1S)(\pi^+\pi^-~{\rm and}~\pi^0\pi^0)$, 11.6092 and 5.7875 for $\Upsilon(3S)\to\Upsilon(2S)(\pi^+\pi^-$ ${\rm and}~\pi^0\pi^0)$, respectively. Parameters of the coupling functions of the decay particles ($J/\psi$, $\psi(2S)$, $\Upsilon(2S)$ and $\Upsilon(3S)$) to channel~$i$, obtained in the analysis, are: $(\alpha_2,\beta_2)=(0.0843,0.0385)$,
$(\gamma_{10},\gamma_{11},\gamma_{20},\gamma_{21})=(1.1826,1.2798,-1.9393,-0.9808)$,
$(\delta_{10},\delta_{11},\delta_{20},\delta_{21})=$($-0.1270$,~16.621,~5.983,~$-57.653$),
$(\rho_{10},\rho_{11},\rho_{20},\rho_{21})=$(0.4050, 47.0963, 1.3352,$-21.4343)$,
$(\omega_{10},\omega_{11},\omega_{20},\omega_{21})=$($1.1619,-2.915$,$0.7841$,~1.0179),
$(\tau_{10},\tau_{11},\tau_{20},\tau_{21})=(7.2842,-2.5599,0.0,0.0)$.

Satisfactory combined description of all considered processes is obtained with the total $\chi^2/\mbox{ndf}=596.706/(527-78)\approx1.33$;\\ for the $\pi\pi$ scattering, $\chi^2/\mbox{ndf}\approx1.15$;\\ for $\pi\pi\to K\overline{K}$, $\chi^2/\mbox{ndf}\approx1.65$;\\ for $\pi\pi\to\eta\eta$, $\chi^2/\mbox{ndp}\approx0.87$;\\ for decays $J/\psi\to\phi(\pi^+\pi^-, K^+K^-)$, $\chi^2/\mbox{ndp}\approx1.36$;\\ for $\psi(2S)\to J/\psi(\pi^+\pi^-,\pi^0\pi^0)$, $\chi^2/\mbox{ndp}\approx2.43$;\\ for $\Upsilon(2S)\to\Upsilon(1S)(\pi^+\pi^-,\pi^0\pi^0)$, $\chi^2/\mbox{ndp}\approx1.01$;\\
for $\Upsilon(3S)\to\Upsilon(1S)(\pi^+\pi^-,\pi^0\pi^0)$, $\chi^2/\mbox{ndp}\approx0.67$,\\
for $\Upsilon(3S)\to\Upsilon(2S)(\pi^+\pi^-,\pi^0\pi^0)$, $\chi^2/\mbox{ndp}\approx0.61$.

In Figs. 3 -- 7 we show our fitting to the experimental data on above
indicated decays of $\psi$- and $\Upsilon$-meson families in the combined analysis with the processes
$\pi\pi\!\to\!\pi\pi,K\overline{K},\eta\eta$. Cavities in the energy dependence of di-pion spectra
(Fig. 7, upper panel) is the result of destructive interference between the $\pi\pi$ scattering and
$K\overline{K}\to\pi\pi$ contributions to the final states of decays $\Upsilon(3S)\to\Upsilon(1S)(\pi^+\pi^-,\pi^0\pi^0)$.

\begin{figure}[!thb]
\begin{center}
\includegraphics[width=0.44\textwidth,angle=0]{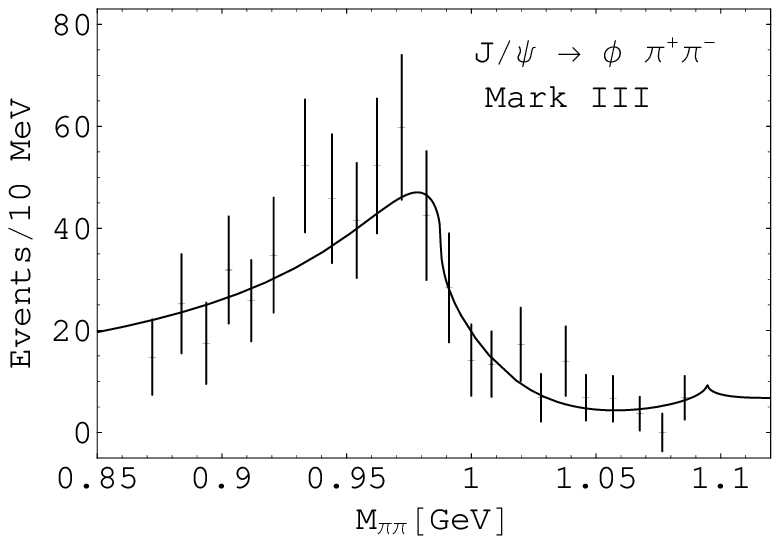}
\includegraphics[width=0.44\textwidth,angle=0]{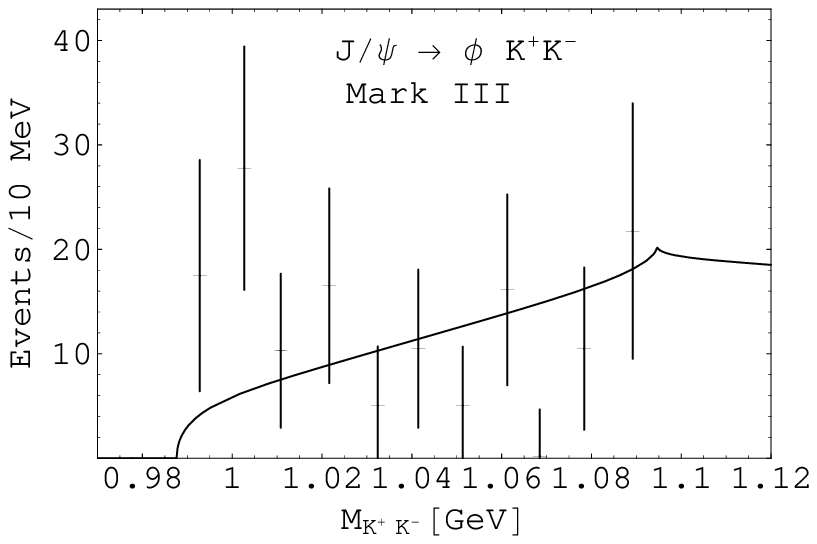}\\
\vspace*{0.1cm}
\includegraphics[width=0.44\textwidth,angle=0]{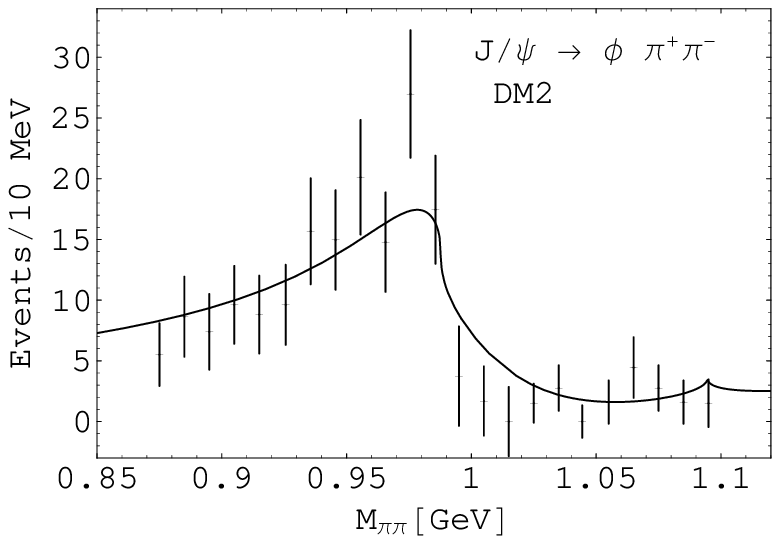}
\includegraphics[width=0.44\textwidth,angle=0]{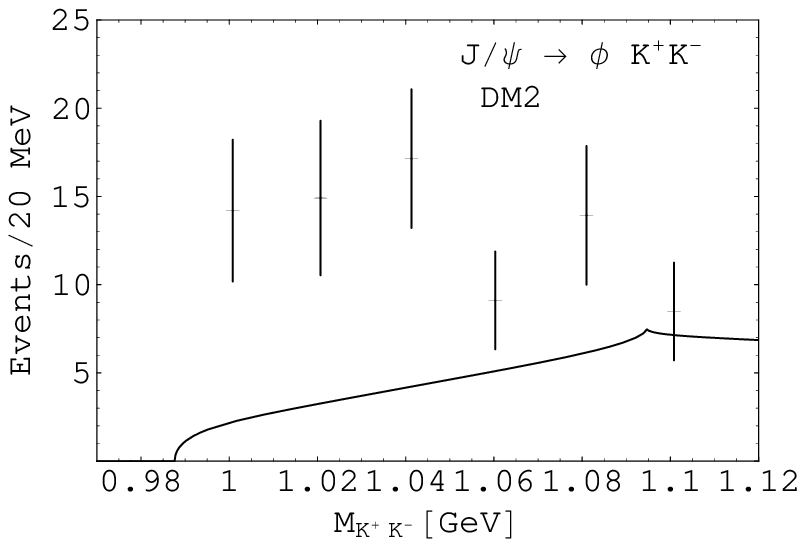}
\vspace*{-0.1cm}\caption{The $J/\psi\to\phi\pi\pi$ and $J/\psi\to\phi
K\overline{K}$ decays. }
\end{center}\label{fig:Jpsi}
\end{figure}

\begin{figure}[!thb]
\begin{center}
\includegraphics[width=0.6\textwidth,angle=0]{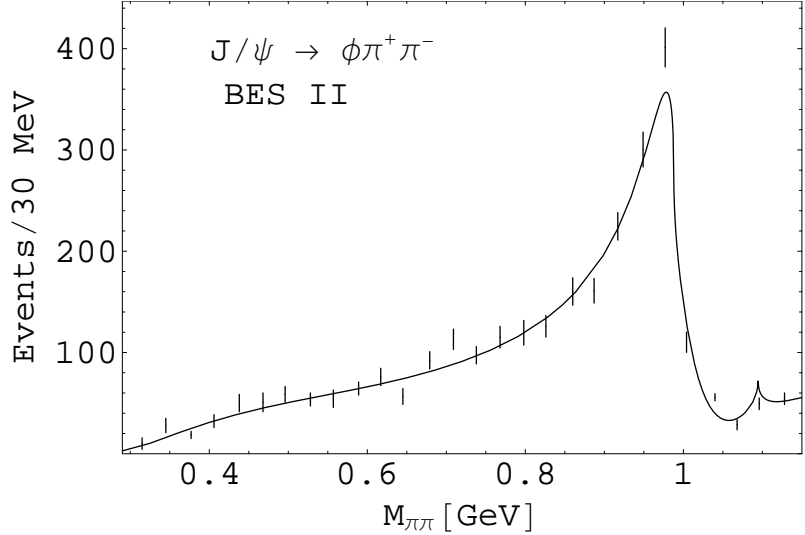}
\vspace*{-0.2cm}\caption{The $J/\psi\to\phi\pi\pi$ decay;
the data of BES~II Collaboration. }
\end{center}\label{fig:BESII}
\end{figure}

Note an important role of the BES~II data:
Namely this di-pion mass distribution rejects the solution with
the narrower $f_0(500)$. The corresponding curve lies considerably below the data from the threshold to about 850~MeV.

\begin{figure}[htb]
\begin{center}
\includegraphics[width=0.45\textwidth]{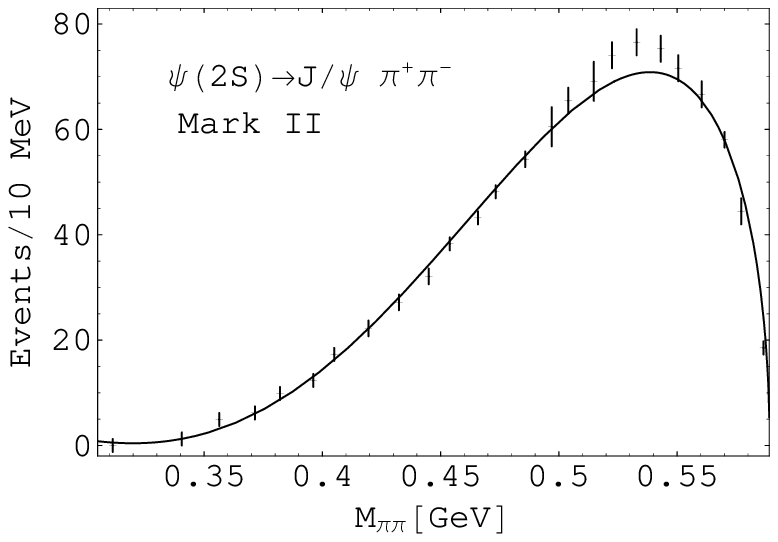}
\includegraphics[width=0.45\textwidth]{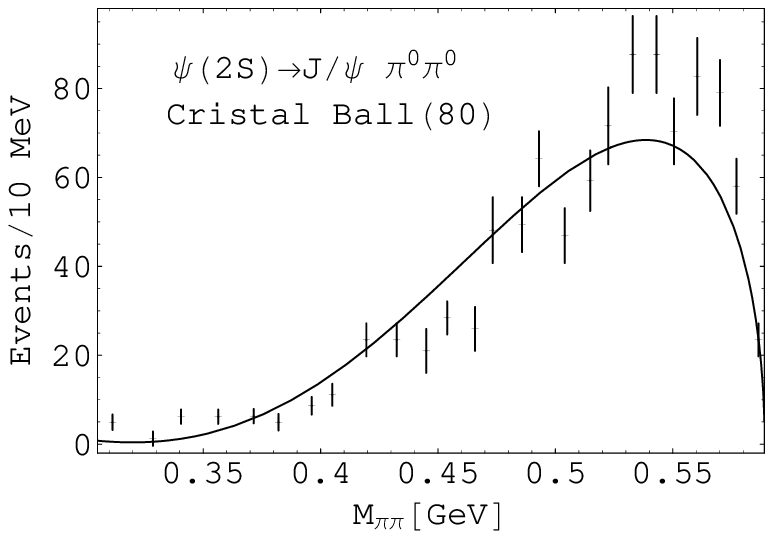}
\end{center}
\vspace*{-.1cm}\caption{The $\psi(2S)\to J/\psi\pi\pi$ decay. }
\label{fig:psi21}
\end{figure}

\begin{figure}[htb]
\begin{center}
\includegraphics[width=0.44\textwidth]{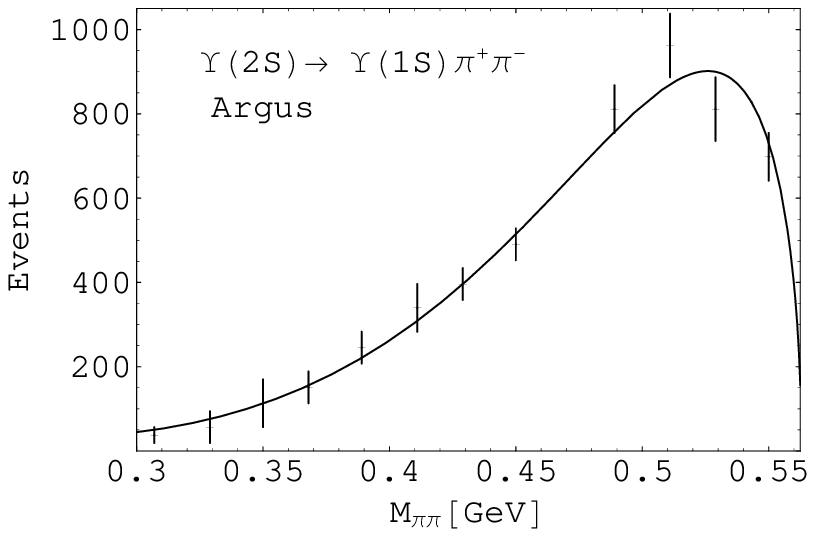}
\includegraphics[width=0.44\textwidth]{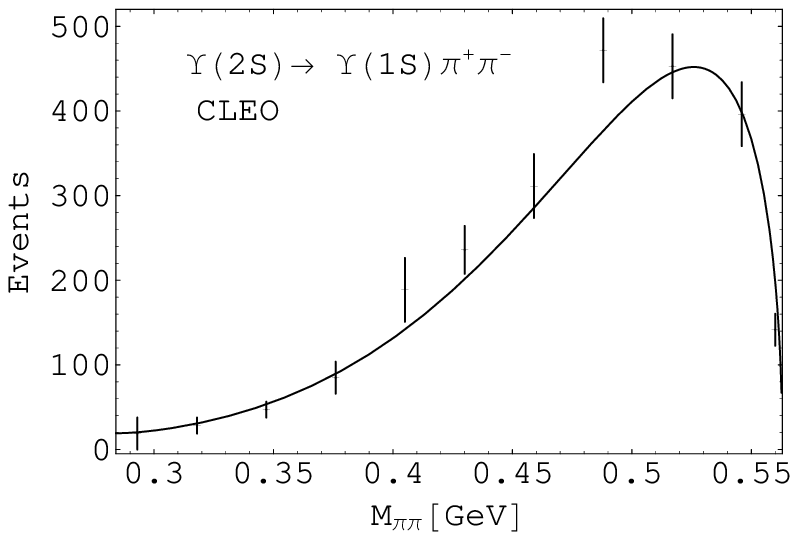}\\
\includegraphics[width=0.44\textwidth]{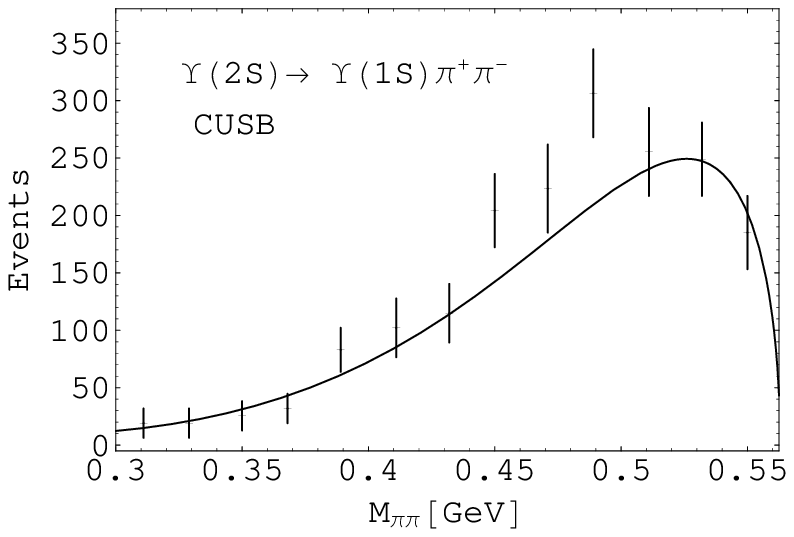}
\includegraphics[width=0.44\textwidth]{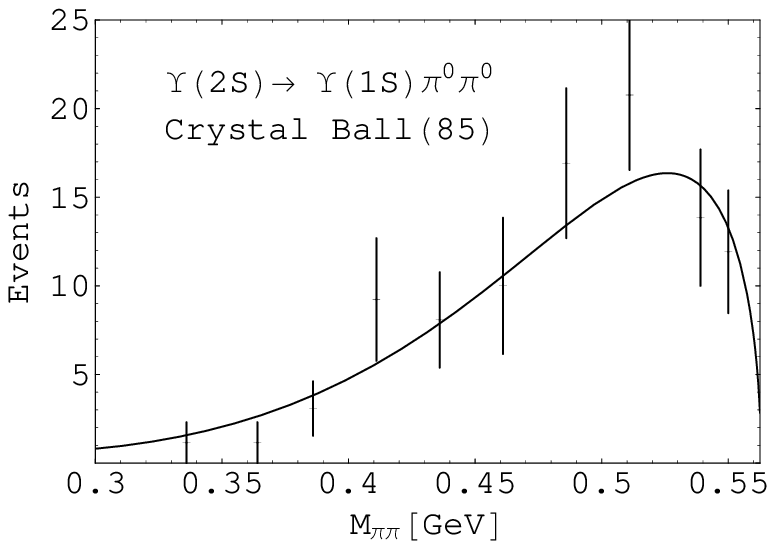}
\vspace*{-0.1cm}\caption{{\small The $\Upsilon(2S)\to\Upsilon(1S)\pi\pi$ decay.
}}
\end{center}\label{fig:Ups21}
\end{figure}

%\vspace*{0.1cm}
\begin{figure}[htb]
\begin{center}
\includegraphics[width=0.44\textwidth]{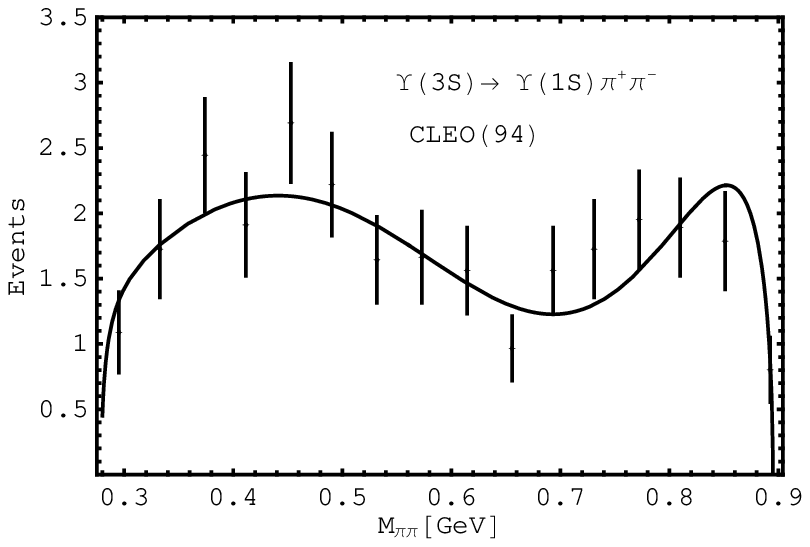}
\includegraphics[width=0.44\textwidth]{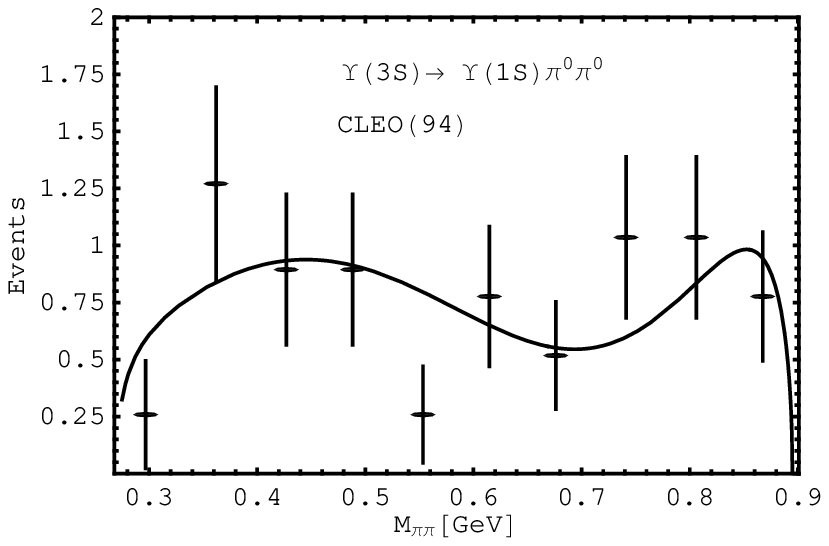}\\
\includegraphics[width=0.44\textwidth]{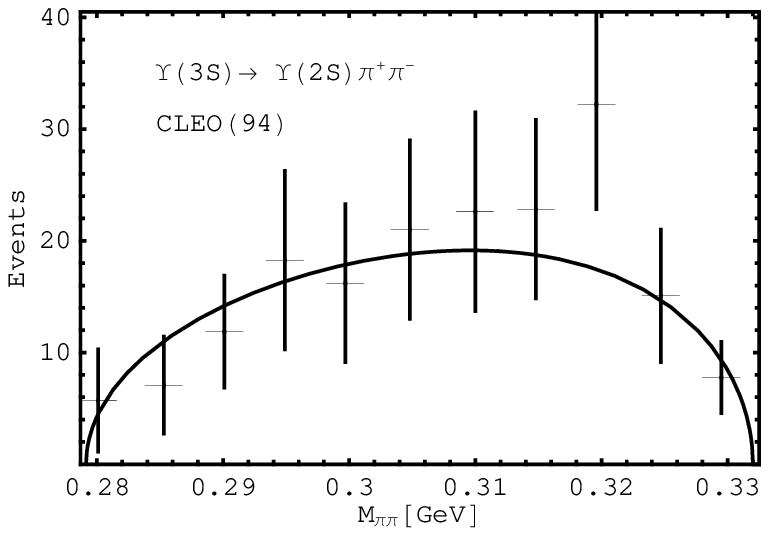}
\includegraphics[width=0.44\textwidth]{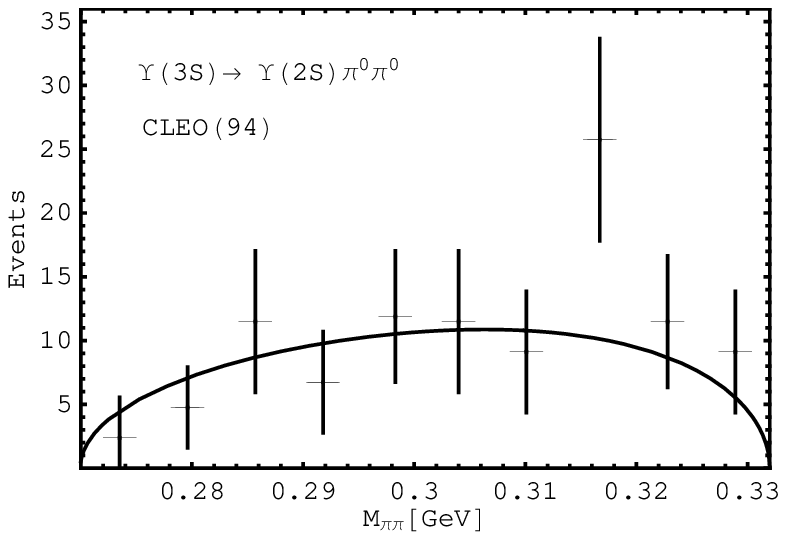}
\vspace*{-0.1cm}\caption{{\small The decays $\Upsilon(3S)\to\Upsilon(1S)\pi\pi$ and $\Upsilon(3S)\to\Upsilon(2S)\pi\pi$.}}
\end{center}\label{fig:Ups31_32}
\end{figure}

\section{Conclusions}

\begin{itemize}

\item
We have performed the combined analysis of data on isoscalar S-wave processes $\pi\pi\to\pi\pi,K\overline{K},\eta\eta$ and on decays
$J/\psi\to\phi(\pi\pi, K\overline{K})$, $\psi(2S)\to J/\psi(\pi\pi)$,
$\Upsilon(2S)\to\Upsilon(1S)\pi\pi$, $\Upsilon(3S)\to\Upsilon(1S)\pi\pi$ and $\Upsilon(3S)\to\Upsilon(2S)\pi\pi$ from the Argus, Crystal Ball,  CLEO, CUSB, DM2, Mark~II, Mark~III, and BES~II Collaborations.

\item
It was shown that in the final states of the $\Upsilon$-meson family decays (except the $\pi\pi$ scattering) the contribution of the coupled processes, e.g., $K\overline{K}\to\pi\pi$, is important even if these processes are energetically forbidden. This is in accordance with our previous conclusions on the wide resonances \cite{SBLKN-prd14,SBLKN-jpgnpp14,SBKLN-PRD12}: If a wide resonance cannot decay into a channel which opens above its mass but the resonance is strongly connected with this channel (e.g. the $f_0(500)$ and the $K\overline{K}$ channel), one should consider this resonance as a multi-channel state with allowing for the indicated channel taking into account the Riemann-surface sheets related to the threshold branch-point of this channel and performing the combined analysis of the considered and coupled channels. E.g., on the basis of that consideration the new and natural mechanism of the destructive interference in the decay $\Upsilon(3S)\to\Upsilon(1S)\pi\pi$ is indicated, which provides the two-humped shape of the di-pion mass distribution (Fig.~7).
\item
Results of the analysis confirm all of our earlier conclusions on the scalar mesons, main of which are:  \\1) Confirmation of the $f_0(500)$ with a mass of about 700~MeV and a width of 930~MeV. This mass value is in line with prediction ($m_{\sigma}\approx m_\rho$) on the basis of mended symmetry by S.Weinberg \cite{Weinberg90} and with an analysis using the large-$N_c$ consistency conditions between the unitarization and resonance saturation suggesting $m_\rho-m_\sigma=O(N_c^{-1})$ \cite{Nieves-Arriola}. Also the prediction of a soft-wall AdS/QCD approach \cite{GLSV_13} for the mass of the lowest $f_0$ meson -- 721~MeV -- practically coincides with the value obtained in our work.\\
2) Indication for the $f_0(980)$ (the pole on sheet~II is $1008.1\pm3.1-i(32.0\pm1.5)$) to be a non-$q{\bar q}$ state, e.g., the bound $\eta\eta$ state. Note that for a earlier popular interpretation of the $f_0(980)$ as a $K\overline{K}$ molecule, it is important whether the mass value of this state is below the $K\overline{K}$ threshold or not.
In the PDG tables of 2010 its mass is 980$\pm$10~MeV. We found in all combined analyses of the multi-channel $\pi\pi$ scattering the $f_0(980)$  slightly above 1~GeV, as in the dispersion-relations analysis only of the $\pi\pi$ scattering \cite{GarciaMKPRE-11}. In the PDG tables of 2012, for the mass of $f_0(980)$ an important alteration appeared: now there is given the estimate 990$\pm$20~MeV.\\
3) Indication for the ${f_0}(1370)$ and $f_0 (1710)$ to have a dominant $s{\bar s}$ component. This is in agreement with a number of experiments \cite{Amsler,Braccini,Barate}.\\
4) Indication for two states in the 1500-MeV region: the $f_0(1500)$ ($m_{res}\approx1495$~MeV, $\Gamma_{tot}\approx124$~MeV) and the $f_0^\prime(1500)$ ($m_{res}\approx1539$~MeV,
$\Gamma_{tot}\approx574$~MeV). The $f_0^\prime(1500)$ is interpreted as a glueball taking into account its biggest width among the enclosing states \cite{Anis97}.

\end{itemize}

\section{Acknowledgments}

This work was supported in part by the Heisenberg-Landau Program, the Votruba-Blokhintsev Program for Cooperation of Czech Republic with JINR, the Grant Agency of the Czech Republic (grant No. P203/12/2126), the Bogoliubov-Infeld Program for Cooperation of Poland with JINR, the DFG under Contract No. LY 114/2-1, the Tomsk State University Competitiveness Improvement Program, and by the Polish National Science Center (NCN) grant DEC-2013/09/B/ST2/04382.

\end{document}